\documentclass[11pt]{article}
\usepackage{moriond,epsfig}
\usepackage{amsmath,amssymb}

\def\be{\begin{equation}}
\def\ee{\end{equation}}
\def\bea{\begin{eqnarray}}
\def\eea{\end{eqnarray}}

\newcommand{\figref}[1]{Figure \ref{#1}}
\newcommand{\tabref}[1]{Table \ref{#1}}

\renewcommand{\eqref}[1]{(\ref{#1})}

\newcommand{\mslash}[1]{#1\hspace{-1.5ex}\slash\hspace{0.3ex}}

\newcommand{\ptmiss}{\mslash{P}_T}
\newcommand{\pt}{P_T}

\newcommand{\pointAc}{P1}
\newcommand{\pointAa}{P2}
\newcommand{\pointAb}{P3}
\newcommand{\pointB}{P4}

\begin{document}
\vspace*{4cm}
\title{Observing Higgsless LSPs at the LHC}

\author{Alexander Knochel}

\address{Physikalisches Institut, Universit\"at Freiburg, Hermann-Herder-Str. 3a \\
79104 Freiburg, Germany}

\maketitle\abstracts{
  We explore the LHC phenomenology of the dark matter candidate arising from the extension of warped Higgsless models. In particular, we consider a model of warped supersymmetry in the bulk and on the IR brane in which the lightest Neutralino is rendered stable by an R parity and serves as a realistic cold dark matter candidate. The production of the LSP and NLSP in association with third generation quarks is simulated using an implementation in {\it O'Mega/Whizard}.
}
\section{The Model}
\label{sect_model}
The model \cite{Knochel:2008ks} is based on Higgsless models proposed in \cite{Csaki:2003dt,Cacciapaglia:2004rb}. They are constructed in 5D using a warped background inspired by the RSI scenario \cite{Randall:1999ee},
$ g_{\mu\nu}=\eta_{\mu\nu}e^{-2 R k y}\quad g_{55}=-R^2\quad y \in [0,\pi]$.
The 5D gauge group
$G_{bulk}=SU(3)_C\times SU(2)_L \times SU(2)_R \times U(1)_{B-L}$
is broken by boundary conditions to
$G_{SM}=SU(3)_C\times SU(2)_L\times U(1)_Y$ on the UV brane and
$SU(3)_C\times SU(2)_D\times U(1)_{B-L}$ on the IR brane.
and thus only $SU(3)_C\times U(1)_{EM}$ survives as a conserved subgroup.
There are at least eight real supercharges in 5D relating it to 4D $\mathcal{N}=2$ SUSY, but half of the symmetries are already broken by the background, leaving us with usual $\mathcal{N}=1$ SUSY after Kaluza-Klein expansion. The action of 5D SYM theory broken by a warped background can be written down using $\mathcal{N}=1$ superfields $A_\mu, \lambda_1,D \in V$ (vector), $A_5,\Sigma,\lambda_2,F_V \in \chi$ (chiral), 
while the hypermultiplet can be written down using one chiral and one antichiral superfield $H, \overline{H}^c$. The complete bulk superfield content is given in \tabref{fig_sfields}.
\begin{table}[b]
\caption{The superfield content of the model and corresponding representations and quantum numbers with respect to the bulk gauge group $G_{bulk}=SU(3)_C\times SU(2)_L \times SU(2)_R\times U(1)_{B-L}$. \label{fig_sfields}}\vspace{0.4cm}
\begin{center}
\begin{tabular}{|l|l|l|l|}
\hline
Superfield& Rep. $G_{bulk}$& Superfield & Rep. $G_{bulk}$\\
\hline\hline
 $V^{Ca}$, $\chi^{Ca}$ &${\bf 8}$ of $SU(3)_C$ & 
 $V^{Li}$, $\chi^{Li}$ &$\bf 3$ of $SU(2)_L$ \\ 
 $V^{Ri}$, $\chi^{Ri}$ & $\bf 3$ of $SU(2)_R$ &
 $V^X$, $\chi^X$ & $U(1)_{B-L}$
\\[0ex] 
 $H^{L}_{l,g}$&$(\mathbf{1},\mathbf{2},\mathbf{1},-1)$ &
 $H^{R}_{l,g}$&$(\mathbf{1},\mathbf{1},\mathbf{2},-1)$ \\ 
 $H^{Lc}_{l,g}$&$(\mathbf{1},\mathbf{\overline{2}},\mathbf{1},1)$ &
 $H^{Rc}_{l,g}$&$(\mathbf{1},\mathbf{1},\mathbf{\overline{2}},1)$\\[0ex]  
 $H^{L}_{q,g}$&$(\mathbf{3},\mathbf{2},\mathbf{1},1/3)$ &
 $H^{R}_{q,g}$&$(\mathbf{3},\mathbf{1},\mathbf{2},1/3)$  \\ 
 $H^{Lc}_{q,g}$&$(\mathbf{\overline{3}},\mathbf{\overline{2}},\mathbf{1},-1/3)$ &
 $H^{Rc}_{q,g}$&$(\mathbf{\overline{3}},\mathbf{1},\mathbf{\overline{2}},-1/3)$\\ \hline
\end{tabular}
\end{center}
\end{table}
To obtain the spectrum of the model, we still have to assign boundary conditions. The IR brane (i.\,e.~$y=\pi$) boundary conditions are a straightforward generalization of the nonsupersymmetric boundary conditions \cite{Knochel:2008ks},
	\begin{subequations}
\label{superbcs}
\begin{align}
  \begin{bmatrix} 1 & -1  \\ \partial_y & \partial_y  \end{bmatrix}
  \left.\begin{bmatrix} V^L \\ V^R \end{bmatrix}\right|_{y=\pi}
  =
  \left.\begin{bmatrix}
    \partial_y & -\partial_y \\
    1 & 1
  \end{bmatrix} e^{-2 R k y}
  \begin{bmatrix} \chi^L \\ \chi^R \end{bmatrix}\right|_{y=\pi} &= 0\,,\\
  \partial_y  V^X(\pi)
     = \chi^X(\pi)
     = \partial_y V^C(\pi)
     = \chi^C(\pi)
    &= 0,\\
  H^{Lc}_{g}(\pi)+\mu_g H^{Rc}_{g}(\pi)=H^R_{g}(\pi)-\mu_g H^L_{g}(\pi)&=0
\end{align}
\end{subequations}
where $g=l_1\dots l_3, q_1\dots q_3$ runs over all leptons and quark generations, and $\mu_g \Lambda_{IR}$ is the IR Dirac boundary mass parameter of the $g$th doublet. As a contrast, the UV brane does not carry SUSY which means that boundary conditions can differ within 4D multiplets.
The physical scalars $h_f$, $h^c_f$, $\Sigma^i$ thus get universal Dirichlet conditions, while the gauginos receive twisted and mixing boundary conditions,
\begin{subequations}
\begin{align}
h_f(0)=h^c_f(0)=\Sigma^i(0)=\lambda_1^C(0)=\lambda^{L}_1
   = \lambda^{R12}_2 =0\\
      \cos(\theta_N) \lambda^X_1+\sin(\theta_N) \lambda^{R3}_1
   = \cos(\theta_N) \lambda^{R3}_2-\sin(\theta_N) \lambda^X_2
    = 0
\end{align}
\end{subequations}
While the parameter space of the model is still rather large, there are several reasonable assumptions that one can make. First of all, we impose (tree level) degeneracy of the pairs of electroweak gaugino modes which will be lifted only at the loop level (no Majorana masses on the UV brane). Furthermore, the splitting of the $W$ and $\chi^+$ raises the KK scale and is therefore assumed to be small (with the lightest charginos just above the experimental lower bounds). The neutralino mixing angle $\theta_N$ is then fixed by the relic density \cite{Knochel:2008ks}. The localization of the matter hypermultiplets is controlled by the multiplet bulk mass $c=M_5/k$. The localization of the light quarks is largely determined by the S parameter to be around $c_L \approx 0.5$ which also suppresses the coupling to the heavy resonances. The third generation is naturally IR localized to generate the heavy top. While this basically fixes the properties which are relevant to the LSP production processes which we will discuss later, there is some freedom in these localization parameters which strongly impact LHC phenomenology.
Exactly delocalized light quarks ($c_L=-c_R=1/2$) have vanishing couplings to KK gluons, while a small deviation from delocalization introduces nonzero couplings.
At the same time, localized kinetic quark terms as they are used to split the doublets introduce a localization effect on the UV brane which also shifts these effective couplings to a nonzero value. Depending on the exact choices, the production of KK gluons is irrelevant or observable in our study of LSP production at the LHC. 
Our minimal implementation of the third generation, though not addressing the $Zbb$ problem, provides a simple way to study the phenomenology of the $t$ and $b$ in LSP production for different scenarios from strongly IR localized fields to the almost delocalized case. 
The introduction of a UV localized kinetic term for the quarks shifts the effective localization. While the localization of the third generation lets the mass of the first quark KK modes vary between extremely light (for almost delocalized third generation fields requiring large IR Dirac masses) and heavy ($\approx 3 k e^{- R k \pi}$), the masses of the lightest $\tilde{t}$ and $\tilde{b}$ modes stay below $2 k e^{- R k \pi}\approx 1100$ GeV and can thus be pair produced at LHC energies with appreciable cross sections.

\section{Production of Missing Energy and heavy Quarks}
\begin{figure}
\begin{center}
\hspace{-3ex}\includegraphics[width=7cm]{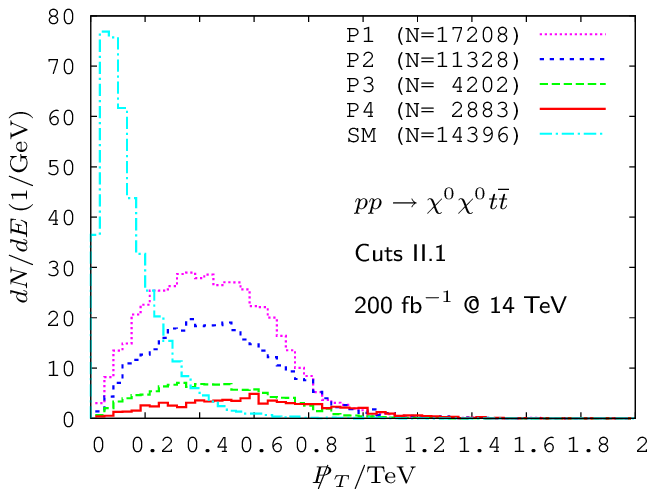}\hspace{-1ex}\includegraphics[width=7cm]{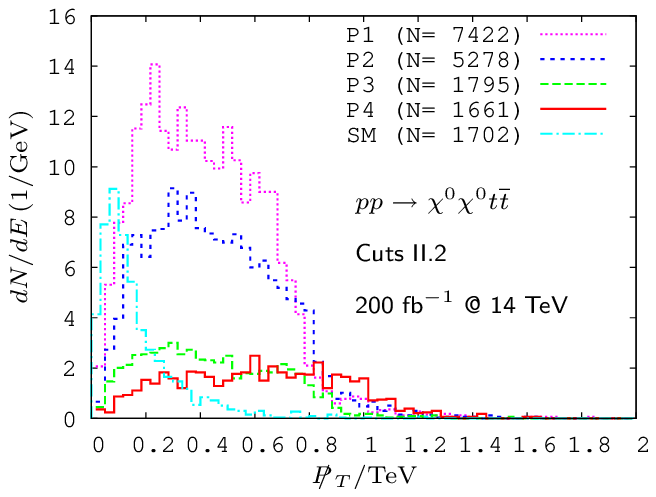}\\
\hspace{-3ex}\includegraphics[width=7cm]{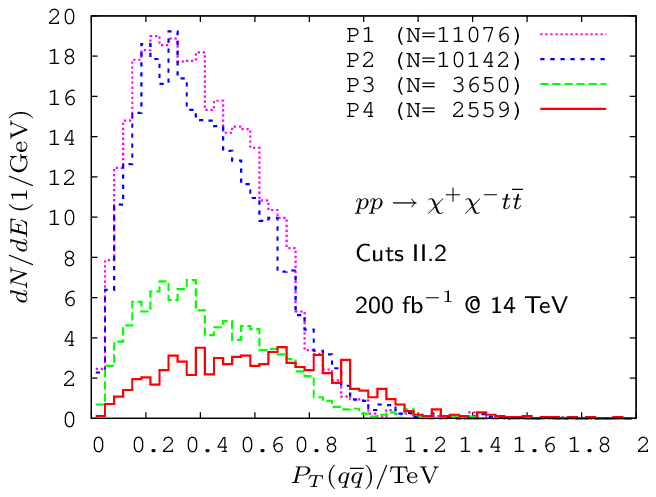}\hspace{-1ex}\includegraphics[width=7cm]{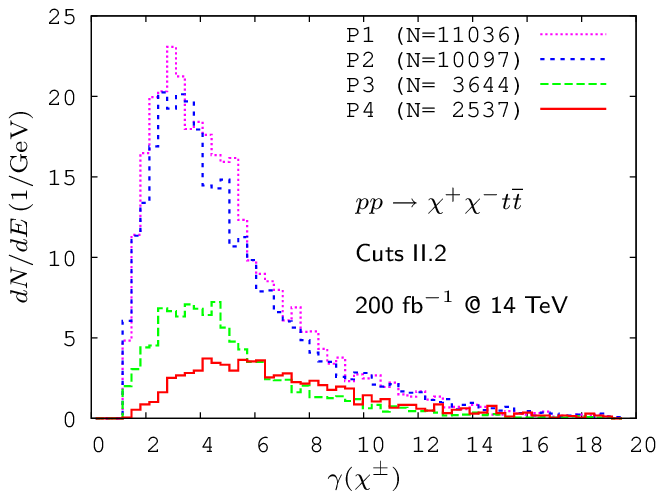}\\
\hspace{-3ex}\includegraphics[width=7cm]{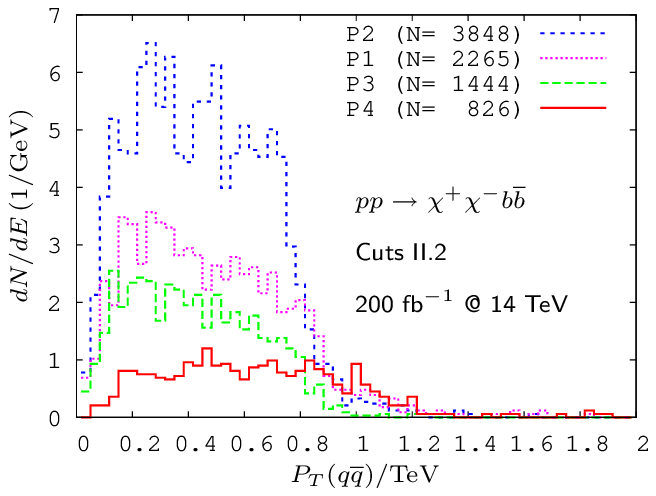}\hspace{-1ex}\includegraphics[width=7cm]{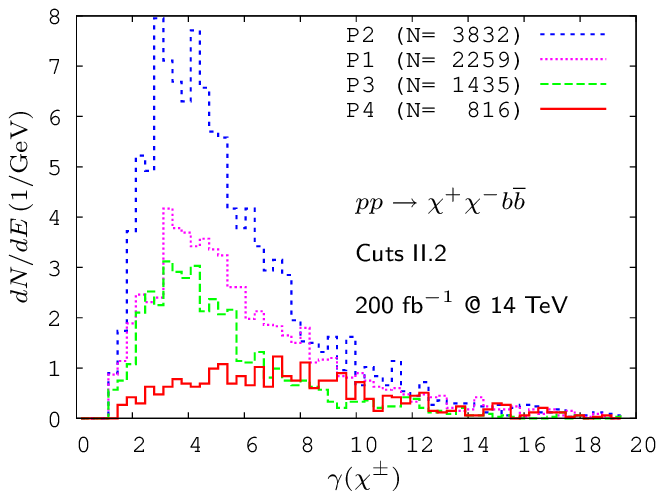}
\end{center}
\caption{
Missing energy from LSP and neutrino production in
  association with top pairs for different quark localizations and
  cuts on invariant masses and azimuthal angle. The line marked SM
  shows the missing energy in $pp\rightarrow \nu \overline{\nu} t
  \overline{t}$.  Below are total transverse momenta and boosts of charginos produced in
  association with top and bottom pairs for different quark
  localizations after cuts II.2. The total $\pt$ is shown as an
  approximation to $\ptmiss$ which will have further contributions
  from the decay products (all MSTW08).\label{fig_chapt}\label{missingptsummary}}
\end{figure}

We concentrate on a set of final states which is particularly favored in the model which we consider in this work, the production of third generation quarks in association with missing energy. 
In our scenario, the first stop mode $\tilde{t}$ is in a convenient mass range: it is still light enough to be pair produced copiously at the LHC at 14 TeV, and at the same time heavy enough ($m_{\tilde{t}}-m_{\chi}-m_t\gtrsim 400$ GeV) in all but the extreme cases to produce a strong missing energy signal from the decay. Such a situation has been discussed in a generic way in \cite{Han:2008gy}. 
The analysis carried out by the authors is valid for our $\tilde{t}$ pair production contributions, but this is only one of the contributions to this class of final states in our model, where the production of heavy quark and gluon resonances proves to be important as well. Due to the size of the model, we rely on simulations with four particle final states, which means that we do not consider the possible decay modes of the $t$ (hadronic, semileptonic, leptonic).
We consider a set of points in parameter space representing different localizations of light and heavy quarks (\tabref{tab:Pn}), while the IR scale is assumed to be $\Lambda_{IR}\approx 620$ GeV and $m_{\chi0}=88$ GeV, $m_{\chi+}=103$ GeV. To proceed, we define a number of cuts
\begin{table}
\caption{
Points in bulk mass parameter
  space of the first and second ($c_{1,2}$) and third ($c_3$)
  generation of quarks.\label{tab:Pn}\label{tabcuts}}\vspace{0.4cm}
\begin{center}
\begin{tabular}{|c|c|c|c|c|}\hline
Bulk Mass&\pointAc{}&\pointAa{}&\pointAb{}&\pointB{}\\
\hline\hline $c_{L1,2}$ &0.48&0.48&1/2&0.48 \\
\hline $c_{R1,2}$ &-0.48&-0.48&-1/2&-0.48 \\
\hline $c_{L3}$ &1/3&0.4&0.4& 0.2\\
\hline $c_{R3}$ &-0.4&-1/3&-1/3& -0.2\\ \hline
\end{tabular}
\end{center}
\end{table}
in addition to the standard cuts 
$M(q,\overline{q})\in[10,\infty]$;
$M(\mbox{parton},q),M(\mbox{parton},\overline{q})\in[-\infty,-10]$;
$E(\mbox{parton})>20$;
$\eta(q),\eta(\overline{q}) \in [-5,5]$ to further suppress backgrounds. The ones used here
are $\pt{(q)},\pt{(\overline q)}>100 \mbox{ GeV}$ (II.1) and  $\pt{(q)},\pt{(\overline q)}>300 \mbox{ GeV}$ (II.2).
When judging the results for the missing energy signal with $t$ and $b$ quarks in the final state, one therefore has to remember the following points: The $t$ pair production itself does not introduce $\ptmiss$, but the leptonic and semileptonic decay modes contain neutrinos, and considering the relative strength of $t$ pair production, this can constitute an important background. In addition to this, there are SM processes with have final states distinguishable from our signal only by their kinematics, for example $pp\rightarrow b\overline{b}\nu l jj$ \cite{Han:2008gy}. While the analysis of these contributions is beyond the scope of this work, there is a generic way to suppress these backgrounds as we can fortunately afford to place rather strong $\pt$ and $\Delta \phi$ cuts without losing too much of our signal.  Some results for the production of missing energy in association with top quarks are shown in \figref{missingptsummary}.  
At the $2\rightarrow 4$ particle level we compare it to the production of missing energy via neutrinos ($pp\rightarrow \nu\overline{\nu}t\overline{t}$), which is the main SM background assuming perfect top quark reconstruction. 
The chargino NLSP in the model discussed here has a mass barely above the current lower bound. As a consequence, it is very narrow ($\Gamma\approx 10^{-7}$ GeV), decaying through an offshell $W$ as $\chi^+ \rightarrow f \overline{f} \chi^0$. Since $\Delta m  \approx 15$ GeV, NLSP production should be visible as missing energy in association with leptons or rather soft jets. As a first approximation, the total transverse momentum of the NLSP pair is shown for cuts II.2 (\figref{fig_chapt}). The production of b pairs in association with missing energy turns out to be not as well suited for this analysis. The SM background in this case is very large, and can only be reduced by severe cuts on transverse momenta, effectively drowning the $pp\rightarrow \chi^0\chi^0 b \overline{b}$ signal.

To conclude, models such as the higgsless supersymmetric scenario investigated in this work, provide an interesting alternative way how electroweak symmetry breaking and dark matter phenomenology can be linked. The discovery of such LSP dark matter candidates at the LHC via production of missing energy in association with top quarks seems promising in large parts of the parameter space. However, it is important to go beyond four particle final states to make more precise statements about the observability of LSP production in this context at the LHC.

\section*{Acknowledgments}
This research was supported by Deutsche Forschungsgemeinschaft through
the Research Training Group 1147 \textit{Theoretical Astrophysics and
Particle Physics}, and by Bundesministerium f\"ur Bildung und
Forschung Germany, grant 05H4\-WWA/2.

\end{document}